\documentclass[english,a4paper,11pt]{article}
\usepackage[margin=3cm]{geometry}
\usepackage[latin1]{inputenc}
\usepackage{latexsym}
\usepackage{booktabs}
\usepackage{amsmath}
\usepackage{algorithm}
\usepackage[final]{graphicx}
\DeclareGraphicsExtensions{.jpg,.jpeg,.pdf,.png,.mps}
\usepackage{epsfig}
\usepackage[round]{natbib}
\usepackage{rotating}
\usepackage{color}
\usepackage{xcolor}
\usepackage{graphicx,subfig}
\usepackage{colortbl}
\usepackage{float}
\usepackage{hyperref}
\usepackage{csquotes}
\usepackage[toc,page]{appendix}
 \usepackage{threeparttable}
\usepackage[bitstream-charter]{mathdesign}
\usepackage[T1]{fontenc}
\usepackage{lmodern}
\title{Enhancing parameter estimation in finite mixture of generalized normal distributions}

\author{\vspace{1.5cm}}

\author{$\mathrm{Pierdomenico \ Duttilo}^\mathrm{1,*},  
	\  \mathrm{Stefano \ Antonio\ Gattone}^\mathrm{2}$\\  
	$^\mathrm{1}$\small{\emph{Department of Statistical Sciences,  University of Padova, Padova, Italy}}\\
    $^\mathrm{2}$\small{\emph{DiSEGS, University "G. d'Annunzio" of Chieti-Pescara, Pescara, Italy}}\\
    $^\mathrm{*}$\small{\emph{Corresponding author: pierdomenico.duttilo@unipd.it}}
}

\date{}

\begin{document}
	\maketitle
\begin{abstract}

\noindent
Mixtures of generalized normal distributions (MGND) have gained popularity for modelling datasets with complex statistical behaviours. However, the estimation of the shape parameter within the maximum likelihood framework is quite complex, presenting the risk of numerical and degeneracy issues.
 This study introduced an expectation conditional maximization algorithm that includes an adaptive step size function within Newton-Raphson updates of the shape parameter and a modified criterion for stopping the EM iterations. Through extensive simulations, the effectiveness of the proposed algorithm in overcoming the limitations of existing approaches, especially in scenarios with high shape parameter values, high parameters overalp and low sample sizes, is shown. A detailed comparative analysis with a mixture of normals and Student-t distributions revealed that the MGND model exhibited superior goodness-of-fit performance when used to fit the density of the returns of $50$ stocks belonging to the Euro Stoxx index. 
\hspace{1cm}\\\\
\textbf{Keywords}: mixture of generalized normal distribution, spurious solutions, ECM algorithm, adaptive step size
\end{abstract}

\section{Introduction}
\label{Intro}
Over time, non-normal mixture distributions have gained increasing attention for modelling random variables with leptokurtic, heavy-tailed, and skewed distributions \citep{Lee2013}. Among the statistical distributions available in the literature, the generalized normal distribution (GND) is capable of describing a large variety of statistical behaviours due to the additional shape parameter that controls the weight of the tails. \citep{Nadarajah2005}. The GND is a natural generalization of the normal distribution, also known as the generalized Gaussian distribution, exponential power distribution, or generalized error distribution. \cite{Nadarajah2005} initially examined the statistical characteristics of the GND, including the hazard rate function, moments, and maximum likelihood estimation, while \cite{Pogany2010} focused on the characteristic function of the GND.

Mixtures of generalized normal distribution (MGND) have attracted considerable interest due to their flexibility to model complex data, including applications in image processing \citep{Bazi2006,Allili2008} and speech modelling \citep{Kokkinakis2005,Deledalle18}. Estimation of parameters is performed using the maximum likelihood estimation (MLE) and the expectation-maximization (EM) algorithm \citep{Dempster1977}. Since the system to resolve the updating equation of the mean and the shape parameters is heavily nonlinear, the numerical optimization based on the Newton-Raphson method is used. However, the Newton-Rapshon update of the shape parameter becomes challenging, especially when its value is greater than 2, causing the algorithm to diverge and produce incorrect values \citep{Roenko2014, Deledalle18}. For example, in the case of the GND distribution, some studies \citep{Kokkinakis2005,KRUPINSKI2006205,Roenko2014,Deledalle18} limit the shape parameter to a range of 0.3-2. As an alternative, \cite{Nguyen2014} proposed a univariate bounded generalised Gaussian mixture model defining a bounded support region in $\mathbb{R}$ for each component of the mixture. \cite{Mohamed2009} estimated the shape parameter using the analytical relationship between the shape parameter and kurtosis, while moment-matching and entropy-matching estimators have been proposed by \cite{Roenko2014} and \cite{Kokkinakis2005}, respectively. Even if these estimators have the advantage of relative simplicity, they are less accurate than MLE. 

Recently, \cite{Wen2020} studied the statistical properties of a two-component MGND and performed MLE using an expectation conditional maximization (ECM) algorithm. The ECM algorithm \citep{Meng1993} is an extension of the EM algorithm which adjusts the M-step into several conditional maximization steps.
\cite{Wen2020} emphasised the need to improve the convergence rate of parameter estimation, suggesting the Regula-Falsi iteration method \citep{Roenko2014} as a potential solution. 

The present work aims to enhance MGND maximum likelihood estimation by including an adaptive step size function for the Newton-Raphson update of the shape parameter and a different stopping criterion for the EM iterations (Algorithm \ref{algo.MGND_GEM} in Appendix \ref{Algorithms}). The proposed algorithm is denoted as ECMs (expectation conditional maximization with step size). 

\begin{figure}[H]
\centering
\includegraphics[width=0.572\textwidth]{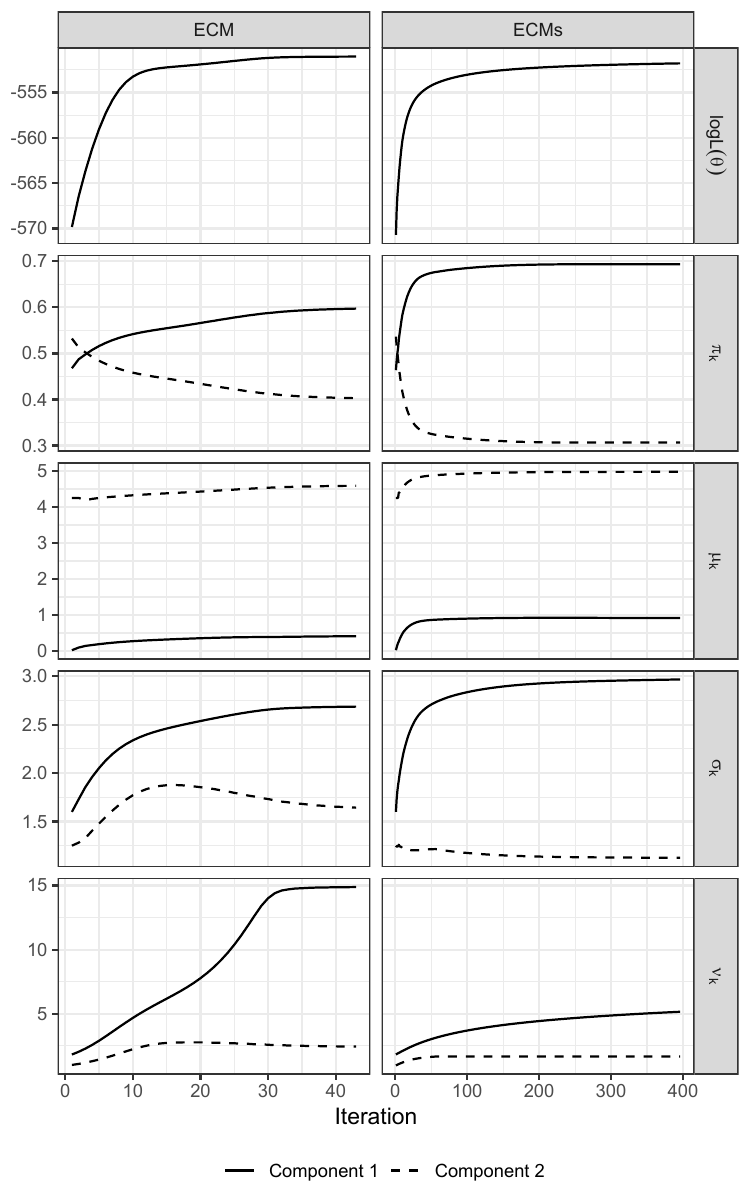}
\caption{Convergence of parameter estimates (motivating example).}\label{fig.convergencepar}	
\end{figure}

\begin{figure}[H]
\centering
\includegraphics[width=1\textwidth]{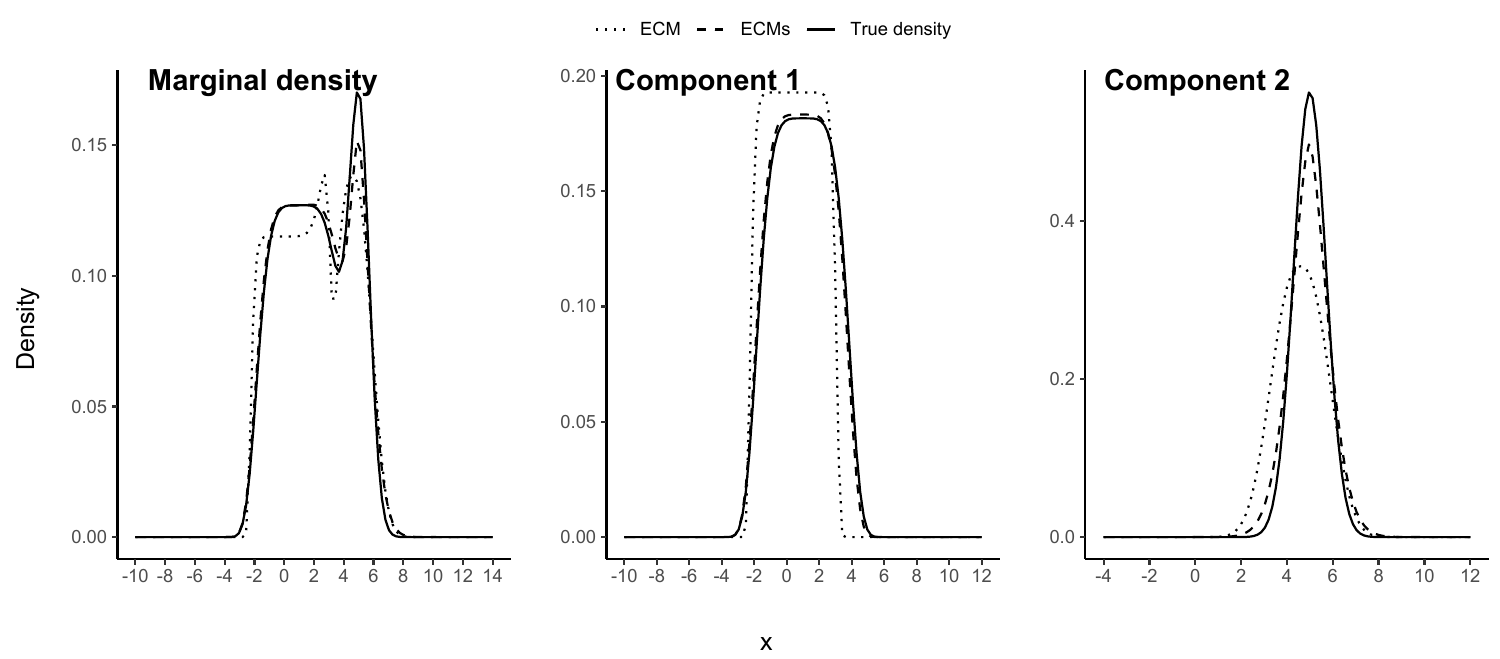}
\caption{Motivating example - estimated densities.}
\label{fig.marginal_comp12}	
\end{figure}

The numerical and degeneracy issues we try to address are illustrated by the following motivating example. Using algorithm \ref{algo.sim} in Appendix \ref{Algorithms}, we simulated a sample of $N=250$ observations from a two-component MGND model with mixture weights $\pi_1=0.7$ and $\pi_2=0.3$, location parameters $\mu_1=1$ and $\mu_2=5$, scale parameters $\sigma_1=3$ and $\sigma_2=1$, and shape parameters $\nu_1=5$ and $\nu_2=1.5$.
Table \ref{cstudy.convergence} reports the log-likelihood and the parameter estimates obtained by applying the plain ECM and the proposed ECMs algorithms together with the estimated variance and kurtosis. 
Figure \ref{fig.convergencepar} shows the convergence of the parameters estimation. With the ECM algorithm, the likelihood function is maximized after $40$ iterations, but the final solution is a spurious solution: the degeneracy of $\widehat{\nu}_1=14.8716$ from its real value $\nu_1=5$ is remarkable. 
On the other hand, the proposed ECMs algorithm avoids the degeneracy of the shape parameter whereas $\widehat{\nu}_1=5.1572$, very close to the true value of the parameter.   
In this example, the degeneracy of $\nu_1$ contributes to worsening the estimates of the other parameters. Figure \ref{fig.marginal_comp12} reports the mixture and component densities estimates obtained with the two algorithms. The ECM algorithm provides a very poor recovery of the single components while obtaining a reasonable estimate of the marginal distribution. However, the fit obtained using the ECMs algorithm is largely improved. 

\begin{table*}[ht]
\caption{Parameter estimates of the ECM and ECMs algorithms (motivating example).}
\label{cstudy.convergence}
\centering
\resizebox{15cm}{!}{
\begin{tabular}[t]{lrrrrrrrrrrccc}
\toprule
&\multicolumn{1}{c}{$\pi_1$}&\multicolumn{1}{c}{$\mu_1$}&
\multicolumn{1}{c}{$\sigma_1$}&\multicolumn{1}{c}{$\nu_1$}&\multicolumn{1}{c}{$\pi_2$}&\multicolumn{1}{c}{$\mu_2$}&\multicolumn{1}{c}{$\sigma_2$}&\multicolumn{1}{c}{$\nu_2$}&\multicolumn{1}{c}{VAR$(X)$}&\multicolumn{1}{c}{Kur$(X)$}&\multicolumn{1}{c}{$\log L(\theta)$}\\
&\multicolumn{1}{c}{$0.7$}&\multicolumn{1}{c}{$1$}& \multicolumn{1}{c}{$3$}&\multicolumn{1}{c}{$5$}&\multicolumn{1}{c}{$0.3$}&\multicolumn{1}{c}{$5$}&\multicolumn{1}{c}{$1$}&\multicolumn{1}{c}{$1.5$}&&&\\
\midrule
ECM &  0.5972 & 0.4100 & 2.6871 & 14.8716 & 0.4028 & 4.589 & 1.6428 & 2.4471 & 
 6.0240 & 1.4011 & -551.0531\\
ECMs &  0.6933 & 0.9149 & 2.9680 & 5.1572 & 0.3067 &  4.9765 & 1.1239 & 1.6666 & 5.7232 & 1.4537 & -551.7986\\
\bottomrule
\end{tabular}
}
\begin{tablenotes}
\item[]{\footnotesize \textit{Notes}. The sample variance and kurtosis are equal to 5.7785 and 1.8966, respectively.}
\end{tablenotes}
\end{table*}

\begin{figure}[H]
 \centering
  \subfloat{\includegraphics[width=.33\linewidth]{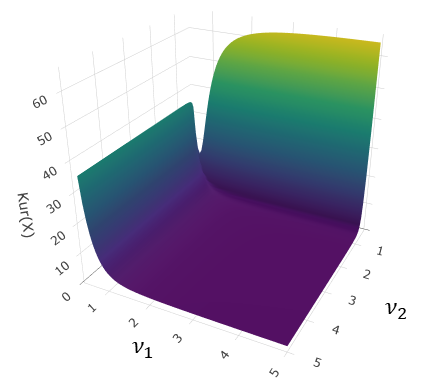}}
  \subfloat{\includegraphics[width=.33\linewidth]{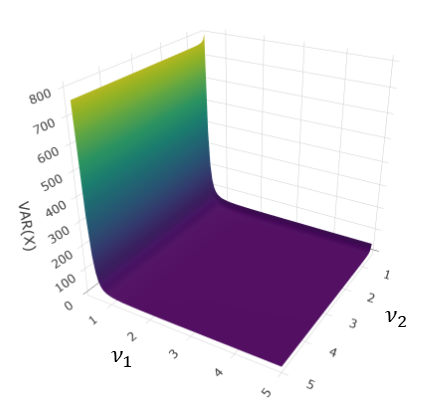}}
  \subfloat{\includegraphics[width=.35\linewidth]{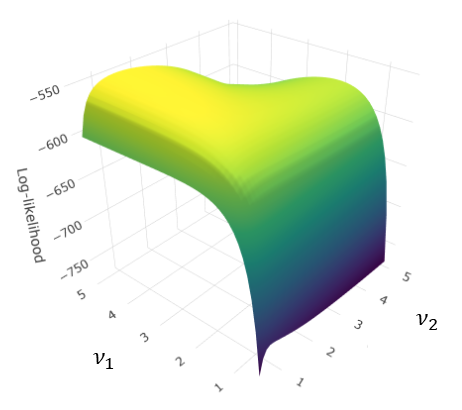}}
  \caption{Motivating example - functional relationships and log-likelihood surface.}
\label{fig.MGNDfr_surface}	
\end{figure}

The poor performance of the ECM algorithm is related to some specific features of the MGND model. From Figure \ref{fig.MGNDfr_surface}, we can observe how the variance, the kurtosis, and the log-likelihood of the MGND model in Table \ref{cstudy.convergence} change as a function of the shape parameters $\nu_1$ and $\nu_2$, keeping all other parameters fixed.
Variance and kurtosis decrease as shape parameters increase but they tend to stabilize over values of $\nu$ greater than 2. In addition, the nearly flat log-likelihood surface that arises when $\nu>2$ (Figure \ref{fig.MGNDfr_surface}) makes it difficult to find sensible estimates, as many values of $\nu$ are fairly equally likely.

Two insights are gained from the motivating example. First, an EM stopping criterion based on the difference of the likelihood value (or the MGND parameters) may not always be a good indicator for stopping the algorithm. Second, a smaller step size could be used in the Newton-Rapshon update to prevent the degeneracy of the shape parameter. In fact, \cite{mythesis} proposed including a step size in the Newton-Raphson update. However, an open issue was the choice of an appropriate value that could vary according to different scenarios.

The rest of the paper is organised as follows. Statistical properties and central moments of the finite mixture of univariate generalized normal distribution are presented in Section \ref{sec.model}. The proposed ECMs algorithm for maximum likelihood estimation (MLE) of the MGND model is presented in Section \ref{sec.parameters} and its performance is evaluated under different simulated scenarios in Section \ref{sec.sim}. An empirical application is presented in Section \ref{sec.data} where the goodness of fit of the two-component MGND model is compared to that of two well-known competing methods: the two-component mixture of normals (MND) and the two-component mixture of Student-t distributions (MSTD). Finally, Section \ref{sec.conclusions} provides some conclusions.

\section{Mixture of generalized normal distribution}\label{sec.model}
A random variable $X$ is said to have the GND with location $\mu$, scale $\sigma$ and shape $\nu$ if its pdf is given by
\begin{equation}
f_{GND}(x|\mu,\sigma,\nu)=\frac{\nu}{2\sigma\Gamma(1\mathbin{/}\nu)}\exp\Biggr\{-\Biggr|\frac{x-\mu}{\sigma}\Biggr|^\nu\Biggr\},
\label{GND}
\end{equation}
with $-\infty<x<\infty$, $-\infty<\mu<\infty$, $\sigma>0$, $\nu>0$, $\Gamma(t)=\int_0^\infty x^{t-1}\exp(-x)dx\text{ for }t>0$.

\begin{figure}[H]
\centering
\includegraphics[width=0.75\textwidth]{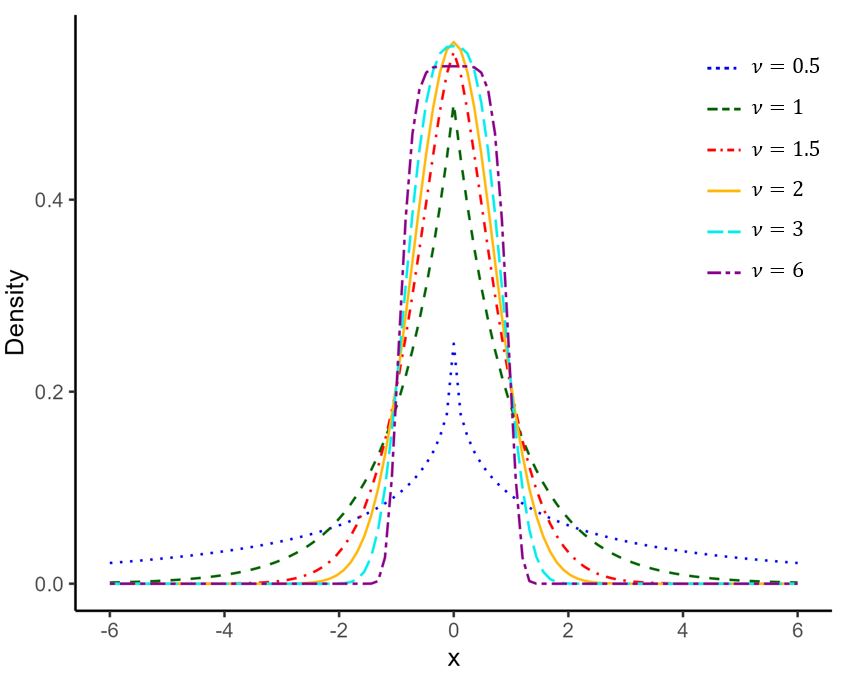}
\caption[GND densities]{GND densities for $\mu=0$, $\sigma=1$ and different shape values.}\label{DensityGND}	
\end{figure}

Figure \ref{DensityGND} shows the probability density function of the GND (Eq. \ref{GND}) for $\mu=1$, $\sigma=1$ and different shape values. The shape parameter $\nu$ controls both the peakedness and the tail weights. If $\nu=1$ the GND reduces to the Laplace distribution and if $\nu=2$ it coincides with the normal distribution. It is noticed that $1<\nu<2$ yields an \enquote{intermediate distribution} between the
normal and the Laplace distribution. As limit cases, for $\nu\rightarrow\infty$ the distribution tends to a uniform
distribution, while for $\nu\rightarrow0$ it will be impulsive \citep{Nadarajah2005,Bazi2006,Dytso2018}. 

The $n$-th central moments, skewness, and kurtosis of the GND are reported in the Appendix (\ref{AGND}).
\cite{Varanasi1989} and \cite{Nadarajah2005} explored two methods for parameter estimation: the method of moments and the MLE. In any case, the estimates do not have a closed form and must be obtained numerically by applying the Newton-Raphson method or the Nelder-Mead approach \citep{Nelder1965}.

A finite mixture of univariate generalized normal distributions (MGND) with $K$ components is given by the marginal distribution of the random variable $X$
\begin{equation}
f_{MGND}(x|\theta)=\sum_{k=1}^K\pi_kf_k(x|\mu_k, \sigma_k, \nu_k),
\label{eq.MGND}
\end{equation}
where $f_k$ is the density $f_{GND}$ of the k-th component and $-\infty<x<\infty$. The set of all the mixture parameters is given by $\theta=\{\pi_k,\mu_k,\sigma_k,\nu_k,k=1,...,K\}$ belonging to the parameter space $\Theta=\{\theta: 0<\pi_k<1, \sum_{k=1}^K \pi_k=1, -\infty<\mu_k<\infty, \sigma_k>0, \nu_k>0, k=1,...,K\}$ with $\text{dim}(\theta)=p$. The two-component MGND ($K=2$) is given by 
\begin{equation}
f_{MGND}(x|\theta)=\pi_1f_1(x|\mu_1, \sigma_1, \nu_1)+\pi_2f_2(x|\mu_2, \sigma_2, \nu_2),\hspace{0.2cm}\text{with }\text{dim}(\theta)=7.
\label{eq.2MGND}
\end{equation}
This model nests several distributions as its sub-models,
namely according to the shape parameter value $\nu_k$. For instance, equation \ref{eq.2MGND} reduces to:
\begin{itemize}
\item Normal mixture model for $\nu_1=\nu_2=2$;
\item Laplace mixture model for $\nu_1=\nu_2=1$;
\item Normal-Laplace mixture model for $\nu_1=2$ and $\nu_2=1$;
\item Normal-GND mixture model for $\nu_1=2$ and $\nu_2>0$;
\item Laplace-GND mixture model for $\nu_1=1$ and $\nu_2>0$.
\end{itemize}

Looking at the $n$ -th central moments, the skewness and kurtosis of the MGND reported in the appendix (\ref{AMGND}), it is clear that these quantities depend on all the parameters of the mixture and, as already seen in Figure \ref{fig.MGNDfr_surface}, variance and kurtosis decrease as $\nu_1$ and $\nu_2$ increase.

\section{The ECMs algorithm}\label{sec.parameters}

In this section, we modify the ECM algorithm for parameter estimation of the MGND model \citep{Bazi2006, Wen2020} by introducing an adaptive step size function for the Newthon-Raphson update of the shape parameter and by modifying the standard EM stopping criterion based on the difference of the likelihood value.  

A complete description of the ECM algorithm and the proposed ECMs algorithm is provided in Algorithm \ref{algo.MGND_ECM} and Algorithm \ref{algo.MGND_GEM} of Appendix \ref{Algorithms}, respectively.

From equation \ref{eq.2MGND} the log-likelihood function is given by 
\begin{equation}
\log L(\theta)=\sum_{n=1}^N\log{\Biggr[\sum_{k=1}^K\pi_kf_k(x_n|\mu_k, \sigma_k, \nu_k)\Biggr]}.
\label{eq.MGNDloglikelihood}
\end{equation}

The \textbf{E-step} involves computing the Q-function
\begin{equation}
Q(\theta,\theta^{(m-1)})=\sum_{k=1}^K\sum_{n=1}^Nz_{nk}^{(m-1)}\log{\Biggr[\pi_k^{(m-1)}f_k(x_n|\mu_k^{(m-1)}, \sigma_k^{(m-1)}, \nu_k^{(m-1)})\Biggr]},
\vspace{1cm}
\label{E-step}
\end{equation}
where 
\begin{equation}
z_{nk}^{(m-1)}=\frac{\pi_k^{(m-1)}f_k(x_n|\mu_k^{(m-1)}, \sigma_k^{(m-1)}, \nu_k^{(m-1)})}{\sum_{k=1}^K\pi_k^{(m-1)}f_k(x_n|\mu_k^{(m-1)}, \sigma_k^{(m-1)}, \nu_k^{(m-1)})}.
\label{responsibility}
\end{equation}
The term $z_{nk}^{(m-1)}$ represents the current estimate of the posterior probability in the $m-1$-th iteration, that is, the probability that the observation $n$ belongs to the group $k$ given the parameters of the current component $\theta^{(m-1)}$.  

The conditional \textbf{M-Step} maximizes the Q-function computed in equation (\ref{E-step}) over $\theta$ so to increases the log-likelihood function in equation \ref{eq.MGNDloglikelihood}. 

\paragraph{Mixture weights}
Set $\frac{\partial Q(\theta,\theta^{(m-1)})}{\partial\pi_k}=0$, then
\begin{equation}\label{eq.Mweights}
\pi_k^{(m)}=\frac{\sum_{n=1}^Nz_{kn}^{(m-1)}}{\sum_{k=1}^K\sum_{n=1}^Nz_{kn}^{(m-1)}}.
\end{equation}

\paragraph{Location parameter}
Set \begin{equation}\label{mu_f}
\begin{split}
\frac{\partial Q(\theta,\theta^{(m-1)})}{\partial \mu_k}&=\frac{\nu_k^{(m-1)}}{\bigr(\sigma_k^{(m-1)}\bigr)^{\nu^{(m-1)}_k}}\Biggr(\sum_{x_n\geq\mu_k^{(m-1)}}^Nz_{kn}^{(m-1)}(x_n-\mu_k^{(m-1)})^{\nu_k^{(m-1)}-1}\\
&-\sum_{x_n<\mu_k^{(m-1)}}^Nz_{kn}^{(m-1)}(\mu_k^{(m-1)}-x_n)^{\nu^{(m-1)}_k-1}\Biggr)=0.
\end{split}
\end{equation}
Since equation \ref{mu_f} is non-linear, the iterative Newton-Raphson method is applied as follows:
\begin{equation}\label{eq.mu_ls}
\mu_k^{(m)}=\mu_k^{(m-1)}-\frac{g\bigr(\mu_k^{(m-1)}\bigr)}{g'\bigr(\mu_k^{(m-1)}\bigr)},
\end{equation}
where $g\bigr(\mu_k^{(m-1)}\bigr)$ is the $Q(\theta,\theta^{(m-1)})$ first order derivative with respect to $\mu_k$, while $g'\bigr(\mu_k^{(m-1)}\bigr)$ is the second order derivative.  
A detailed version of equation \ref{eq.mu_ls} is derived in Appendix \ref{up.eqs}.

\paragraph{Scale parameter}
Set
\begin{equation*}\label{eq.sigmafpartial0}
\begin{split}
\frac{\partial Q(\theta,\theta^{(m-1)})}{\partial \sigma_k}&=\sum_{n=1}^Nz_{kn}^{(m-1)}\biggr(-\frac{1}{\sigma_k^{(m-1)}}\biggr)+\frac{\nu_k^{(m-1)}}{\bigr(\sigma_k^{(m-1)}\bigr)^{\nu_k^{(m-1)}+1}}\\
&\times\sum_{n=1}^Nz_{kn}^{(m-1)}\bigr|x_n-\mu_k^{(m)}\bigr|^{\nu_k^{(m-1)}}=0.
\end{split}
\end{equation*}
after some calculations, it is possible to obtain the following update \citep{Bazi2006,Wen2020}
\begin{equation}\label{eq.sigmak}
\sigma_k^{(m)}=\Biggr[\frac{\nu_k^{(m-1)}\sum_{n=1}^Nz_{kn}^{(m-1)}|x_n-\mu_k^{(m)}|^{\nu_k^{(m-1)}}}{\sum_{n=1}^Nz_{kn}^{(m-1)}}\Biggr]^{\frac{1}{\nu_k^{(m-1)}}}.
\end{equation}

\paragraph{Shape parameter}
As discussed in Section \ref{Intro}, the estimation of the shape parameter within the MLE framework is quite complex presenting the risk of numerical and degeneracy issues. To overcome these issues we introduce an adaptive step size function which depends on the current value of the shape parameter.

Set
\begin{equation}
\begin{split}
\frac{\partial Q(\theta,\theta^{(m-1)})}{\partial \nu_k}&=\sum_{n=1}^Nz_{kn}^{(m-1)}\frac{1}{\nu_k^{(m-1)}}\Biggr(\frac{1}{\nu_k^{(m-1)}}\Psi\Biggr(\frac{1}{\nu_k^{(m-1)}}\Biggr)+1\Biggr)\\
&-\sum_{n=1}^Nz_{kn}^{(m-1)}\Biggr|\frac{x_n-\mu_k^{(m)}}{\sigma_k^{(m)}}\Biggr|^{\nu_k^{(m-1)}} \log \Biggr|\frac{x_n-\mu_k^{(m)}}{\sigma_k^{(m)}}\Biggr|\Biggr]=0.
\end{split}
\end{equation}
Since the above equation is a non-linear equation, the iterative Newton-Raphson method is applied as follows:
\begin{equation}\label{eq.inversenu_ls}
\nu_k^{(m)}=\nu_k^{(m-1)}-\alpha(\nu^{(m-1)})\frac{g\bigr(\nu_k^{(m-1)}\bigr)}{g'\bigr(\nu_k^{(m-1)}\bigr)},
\end{equation}
where $g\bigl(\nu_k^{(m-1)}\bigr)$ is the first-order derivative of $Q(\theta, \theta^{(m-1)})$ with respect to $\nu_k$, $g'\bigl(\nu_k^{(m-1)}\bigr)$ is the second-order derivative and  
\begin{equation}
\alpha(\nu^{(m-1)})=e^{-\nu_k^{(m-1)}}
\end{equation}
represents the proposed adaptive step size function. The function is conceived to avoid the degeneracy of the shape parameter since the magnitude of the Newton step reduces as $\nu_k^{(m-1)}$ increases.

The standard updating equation of the \textit{plain} ECM algorithm \citep{Bazi2006,Wen2020} can be recovered by setting $\alpha(\nu^{(m-1)})=1$. A more detailed derivation of equation \ref{eq.inversenu_ls} is provided in Appendix \ref{up.eqs}.

\paragraph{Stopping criterion}
In the motivating example of Section \ref{Intro} it has been noticed that maximizing the likelihood when updating the shape parameter may lead to spurious solutions. To mitigate this issue, we propose to assess the convergence of the shape parameter by monitoring the first derivative of the Q function with respect to $\nu_k$ and not to update the shape parameter when $g\bigl(\nu_k\bigr)$ is less than a given constant $\eta$. The idea is that $g\bigl(\nu_k\bigr)<\eta$ could be evidence that the likelihood surface is flat with respect to the shape parameter (see Figure \ref{fig.MGNDfr_surface}). In this circumstance, the iterative update of $\nu$ such that the likelihood value increases may lead to a spurious solution. As a backup check, the convergence of the location and scale parameters is evaluated following the standard EM stopping criterion based on the difference in the likelihood value. However, from the simulation study it turned out that once $g\bigl(\nu_k\bigr)<\eta$, the mixing proportions, the mean and the scale parameters have already converged showing that the shape parameters are the last to stabilise.

\section{Simulations}\label{sec.sim}
The main objective of this section is to study the performance of the proposed ECMs algorithm. By comparing across different simulated scenarios the ECMs results with those obtained using the {\it plain} ECM algorithm, we can investigate the benefits of the introduced innovations, {\it i.e.} the adaptive step size function and the modified stopping criterion. The simulated scenarios are shown in Table \ref{tab.MGNDsim_scenarios}. Scenario 1 shares the parameter setting of the motivating example. In Scenario 2, the overlap between the two components of the mixture increases by imposing common location parameters ($\mu_1 = \mu_2 = 0$), while the scale parameters are switched between the components compared to scenario 1. Scenarios 3 and 4 feature a normally distributed main component ($\nu_1 = 2$) and a less frequent component with heavy tails ($\nu_2 = 0.8 < 1$). In Scenario 4, the mixture components have common location parameters.

\begin{table}[H]
\caption{MGND simulated scenarios}
\label{tab.MGNDsim_scenarios}
\centering
\begin{tabular}{ccccccccc}
\toprule
&\multicolumn{1}{c}{$\pi_1$}&\multicolumn{1}{c}{$\mu_1$}&
\multicolumn{1}{c}{$\sigma_1$}&\multicolumn{1}{c}{$\nu_1$}&\multicolumn{1}{c}{$\pi_2$}&\multicolumn{1}{c}{$\mu_2$}&\multicolumn{1}{c}{$\sigma_2$}&\multicolumn{1}{c}{$\nu_2$}\\
\midrule
Scenario 1 & 0.7 & 1 & 3 & 5 & 0.3 & 5 & 1 & 1.5\\  
Scenario 2 & 0.7 & 0 & 1 & 5 & 0.3 & 0 & 3 & 1.5\\ 
Scenario 3 & 0.7 & 1 & 1 & 2 & 0.3 & 5 & 3 & 0.8\\
Scenario 4 & 0.7 & 0 & 1 & 2 & 0.3 & 0 & 3 & 0.8\\
\bottomrule
\end{tabular}
\end{table}
 
For each scenario, $S=250$ samples  are generated with the composition method \citep{rizzo2019} described in Algorithm \ref{algo.sim} (Appendix \ref{Algorithms}) with size $N=250,1000$. In each sample, we run the ECM and ECMs algorithms using the same 10 starting points as described in the Appendix \ref{Algorithms}.  As measures of quality estimation, the average of the estimates (AVG) and the root mean square error (RMSE) are computed for each parameter as follows:
\begin{equation}
\text{AVG}_\theta=\frac{1}{S}\sum_{s=1}^S\widehat{\theta}.
\end{equation}
and
\begin{equation}
\text{RMSE}_\theta=\sqrt{\frac{1}{S}\sum_{s=1}^S(\widehat{\theta}_s-\theta)^2},
\end{equation}
where $\theta$ is the true parameter value and $\widehat{\theta}_s$ is the estimate of $\theta$ for the $s$-th simulated data.

\begin{figure}[H]
\centering
\includegraphics[width=0.90\textwidth]{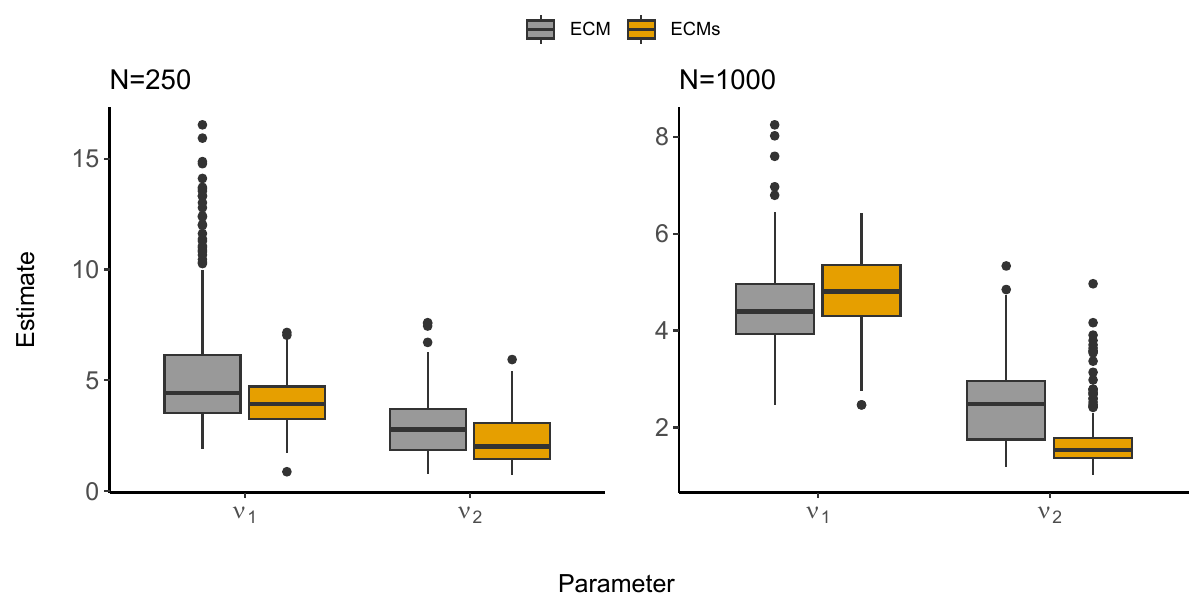}
\caption{Boxplot of shape parameter estimates for Scenario 1 with $\nu_1=5$ and $\nu_2=1.5$.}
\label{sim1_boxplot}	
\end{figure}

Tables \ref{tab.MGNDsim_set1-2} and \ref{tab.MGNDsim_set3-4} (Appendix \ref{Tables}) report all the simulation results. Figure \ref{sim1_boxplot} presents the box plot estimates of the shape parameters in Scenario 1 where the true values of $\nu_1$ and $\nu_2$ are $5$ and $1.5$, respectively. 

The results confirm the conclusions drawn in the motivating example in Section \ref{Intro}. Looking at the left panel of figure \ref{sim1_boxplot}, it is evident that, when the sample size is small, the ECM algorithm suffers a degeneracy problem providing several spurious solutions characterised by very large estimates of the shape parameter $\nu_1$. On the other hand, the proposed ECMs algorithm largely manages to prevent the degeneracy issue. 
For $N=250$, estimates obtained with the ECM algorithm result in $\text{RMSE}_{\nu_1}=2.99$ compared to $\text{RMSE}_{\nu_1}=1.47$ of the algorithm ECMs (Scenario 1 results of Table \ref{tab.MGNDsim_set1-2}). The proposed ECMs algorithm provides more accurate estimates of the other parameters that reduce, for example, the RMSE of $\nu_2$ from $1.96$ to $1.28$.
For large samples ($N=1000$), the proposed ECMs algorithm is still able to improve on ECM by reducing the RMSE for all parameters. The reduction in RMSE is on the order of $27\%$ and $51\%$ in the estimation of $\nu_1$ and $\nu_2$ and of around $45\%$ in the estimation of the mean parameters. Interestingly enough, RMSE reductions are observed also for the estimation of the scale parameters, even if the updates in the two algorithms coincide. In fact, being the ECM characterised by conditional maximization steps, poor parameter updates of $\mu_k$ and $\nu_k$  will of course affect the update of the scale parameter $\sigma_k$.

Degeneracy issues are also observed in scenario 2 when $N=250$ where $\text{RMSE}_{\nu_1}$ equal to $56.65$ for ECM. The issue is solved by the ECMs algorithm where $\text{RMSE}_{\nu_1}=1.35$. Unlike Scenario 1, for large sample sizes ($N=1000$) the degeneracy problem disappears and the accuracies of ECM and ECMs are similar.

As expected, the ECM algorithm does not report any degeneracy problem in Scenario 3 and 4 since the true shape parameter values are not greater than 2. Generally, in these scenarios, similar estimation results are obtained for the mean and scale parameters, while slightly more accurate parameter estimates are obtained for the shape parameter when the ECMs algorithm is used (Table \ref{tab.MGNDsim_set3-4}). 

Further simulation results are reported in the supplementary materials with the same 4 scenarios but with switched mixing proportions, {\it i.e.} $\pi_1=0.3$ and $\pi_2=0.7$. The higher weight in the second component, characterised by larger variability, has the effect of increasing the components overlap. Again, ECM algorithm suffers from degeneracy issues and notable ECMs RMSE gains are spotted for all parameters in the first two scenarios. However, with reversed proportions, the results obtained by ECM and ECMs within scenarios 3 and 4 are no longer similar to each other. Instead, an overall better performance of ECMs is detected for all parameters and sample sizes. 
In summary, the proposed ECMs algorithm  either provides more accurate parameter estimates compared to the ECM algorithm when the shape parameter is greater than 2, the sample size is low and the components overlap is high or returns similar results in the remaining scenarios.

\newpage
\section{Real data analysis}\label{sec.data}
This section presents an analysis of the Euro Stoxx 50 (SX5E) index over the period 2010-2024. Log-returns are computed, and their distribution is analysed with the MGND, mixture of normals (MND) and mixture of student-t distributions (MSTD). The models' performance is assessed using the Akaike Information Criterion (AIC) and Bayesian Information Criterion (BIC). The comparison is also extended to the 50 constituent stocks of the SX5E.

\subsection{Euro stoxx 50}
Data on daily closing prices of the SX5E have been collected from \cite{yahoo} for the time period January 4, 2010 to September 30, 2024 (3786 observations). The return $r_t$ at time $t$ is defined as follows
\begin{equation}\label{eq.log.ret}
r_t=(\log{P_t}-\log{P_{t-1}})100,
\end{equation}
where $P_t$ and $P_{t-1}$ are the closing prices at time $t$ and $t-1$, respectively. 

\begin{figure}[H]
\centering
\includegraphics[width=1\textwidth]{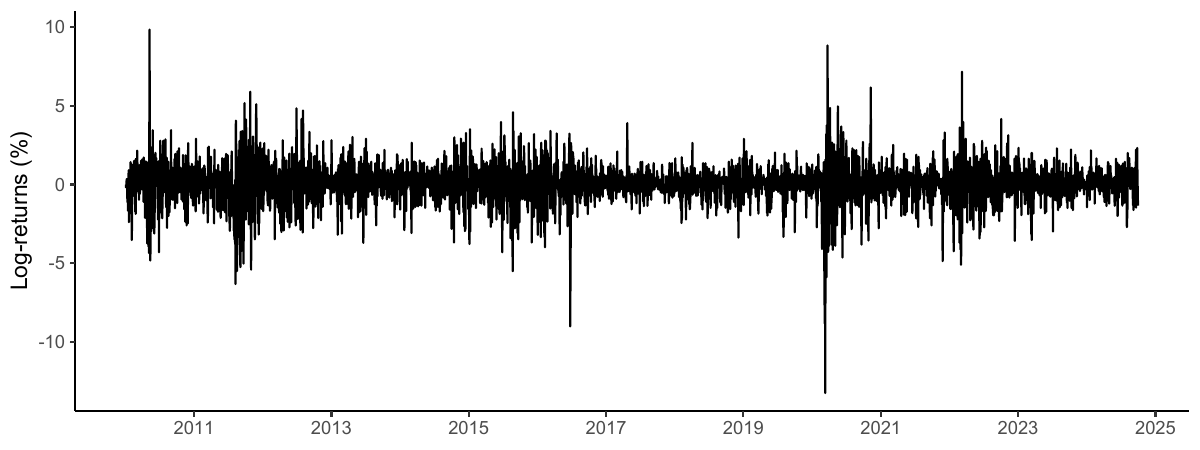}
\caption{Daily returns of the SX5E.}
\label{rt_plot}	
\end{figure}

Figure \ref{rt_plot} shows the time series where two important volatility periods can be observed: the global financial crisis (2007-2008) and the Covid-19 crisis (2020-2021). The daily returns of the SX5E are not normally distributed because they are characterised by heavy tails, leptokurtosis, and negative skewness \citep{Duttilo2021} as shown in table \ref{DesStat}. 

\begin{table}[H]
\begin{center}
	\caption{Descriptive statistic summary.}\label{DesStat}
		\begin{tabular}{lcccccc}
		\toprule
&\multicolumn{1}{c}{Mean}&\multicolumn{1}{c}{Median}&\multicolumn{1}{c}{Std}&\multicolumn{1}{c}{Skew}&\multicolumn{1}{c}{Kur}&\multicolumn{1}{c}{JB Test}\\					
		\midrule
		\textsc{SX5E} & 0.013 &0.042  & 1.265 & -0.466 & 11.0513 & 10377$^{*}$\\
		\bottomrule
		\end{tabular}
		    \begin{tablenotes}
      \item[]{\footnotesize \textit{Notes}. $^{*}$ indicates a \emph{p}-value$\leq0.05$.}
    \end{tablenotes}
  \end{center}
\end{table}

The goodness-of-fit of the two-component MGND model is compared to that of the two-component mixture of normals (MND) and the two-component mixture of Student-t distributions (MSTD). The two-component MGND model is estimated using the proposed algorithm ECMs, while MND and MSTD are estimated with the R package \textit{teigen} \citep{Andrews2011,Andrews2018}. All algorithms employ 5 starting points with k-means initialization. 

Two goodness-of-fit measures are used to find the best-fit model. The Akaike information criterion (AIC) introduced by \cite{Akaike1974} and defined by
\begin{equation}
\text{AIC}=2p-2\log L(\widehat{\theta}),
\end{equation}
where $p$ is the number of parameters of the model and $\log L(\widehat{\theta})$ is the computed log-likelihood. The Bayesian information criterion (BIC) introduced by \cite{Schwarz1978} and defined by
\begin{equation}
\text{BIC}=p\log(N)-2\log L(\widehat{\theta}),
\label{eq.bic}
\end{equation}
where $N$ is the number of observations. Both AIC and BIC balance the goodness-of-fit with model complexity.

The estimation results are reported in Table \ref{SX5Eest}. The best model is the two-component MGND achieving the lowest AIC and BIC values. The MGND model exhibits bimodal asymmetry, with $\mu_1 < \mu_2$. The first component effectively captures extreme market movements through a low shape parameter, $\widehat{\nu}_1 = 0.88841$, while the second component is characterised by a shape parameter of $\widehat{\nu}_2 = 1.2106$, which lies between the normal and Laplace distributions \citep{Duttilo2023}.

\begin{figure}[H]
\centering
\includegraphics[width=0.96\textwidth]{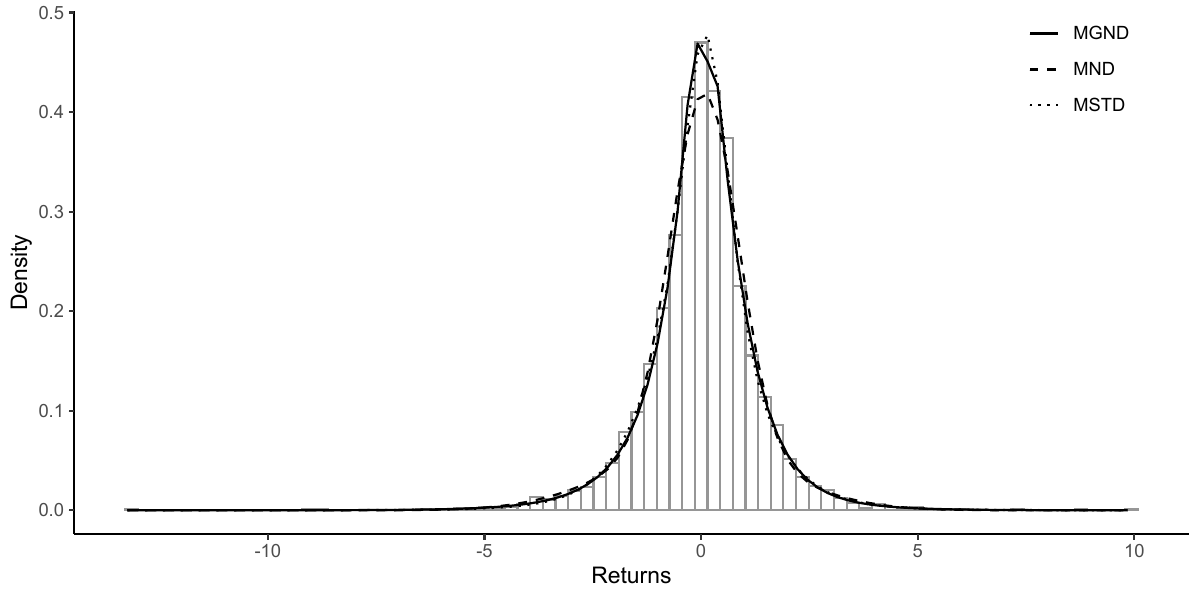}
\caption{Estimated densities for the SX5E.}
\label{rt_densities}	
\end{figure}

Figure \ref{rt_densities} shows the estimated densities of the SX5E. The estimated densities of the MGND and MSTD models are pretty close. 

\begin{table}[H]
\caption{Estimation results for the SX5E.}
\label{SX5Eest}
\centering
\begin{tabular}{lccc}
\toprule
&\multicolumn{1}{c}{MGND}&\multicolumn{1}{c}{MND}&
\multicolumn{1}{c}{MSTD}\\
\midrule
$\pi_1$& 0.5810   & 0.7039& 0.7220\\
$\mu_1$&-0.1779& 0.0793& -0.0058\\
$\sigma_1$& 0.8125& 1.1025& 1.2262\\
$\nu_1$ & 0.8881&-& 4.8312\\
$\pi_2$ & 0.4190 & 0.2961& 0.2780\\
$\mu_2$ & 0.3200 &-0.1334& 0.1239\\
$\sigma_2$ &0.8181& 2.7725& 0.2206\\
$\nu_2$ &1.2106&-& 22.6253\\
\midrule
$\log L(\widehat{\theta})$& -5884.35& -5929.65&-5885.00\\
AIC & \textbf{15058.48}& 15250.81 & 15063.88\\
BIC & \textbf{15104.49}& 15283.67 & 15109.89\\
\bottomrule
\end{tabular}
\centering
\begin{tablenotes}
\item[]{\footnotesize \textit{Notes}. In bold the best model according to the goodness-of-fit measures.}
\end{tablenotes}
\end{table}

\subsection{Comparative analysis on Euro stoxx 50 constituents}
Data on the daily closing prices of the 50 constituents of the SX5E index have been collected from \cite{yahoo} for the period from January 4, 2010, to September 30, 2024. Daily log-returns for each stock are computed according to equation \ref{eq.log.ret}.

Descriptive statistics and the JB test are reported in Table \ref{tab.comp.bs} - Appendix \ref{Tables}. All stocks exhibit a mean and median close to zero. The standard deviation ranges from 1.2370 to 3.1286. Returns are predominantly negatively skewed, with an empirical kurtosis exceeding three, indicating fat-tailed distributions. The JB test rejects the null hypothesis of normality for all stocks.

Tables \ref{tab.comp.aic} and \ref{tab.comp.bic} in Appendix \ref{Tables} report the AIC and BIC values obtained by MGND, MND and MSTD. The AIC equally selects the MGND and MSTD models in 50\% of cases, while the BIC selects the MGND in 50\%, the MND in 2\%, and the MSTD in 48\% of cases. The MGND emerges as a competitive model compared to the MSTD, which is commonly used to fit return distributions \citep{Massing2021}.

\section{Conclusions}\label{sec.conclusions}
The main goal of this paper is to provide a satisfactory algorithm that solves the numerical problem of MLE under a mixture model of generalized normals. Indeed, with generalized normal distributions as the subpopulation distributions, the shape parameter M-step of the EM algorithm does not have an analytical solution. The Newton-Raphson update may be used to solve the maximization problem which however may encounter numerical and degeneracy issues which lead to spurious solutions. By introducing an adaptive step size function and by modifying the EM stopping criterion, we aim at avoiding such spurious solutions. The simulation results have shown that these novelties enhance the accuracy and convergence properties of parameter estimation in MGND models, particularly in scenarios with high shape parameter values, high comopnents overlap and low sample sizes.
In addition, the MGND model is applied to the daily returns of the Euro stoxx 50 and its constituents. Comparison with a two-component mixture of normals (MND) and Student-t distributions (MSTD) revealed that the MGND model exhibited superior goodness of fit performance. The MGND model effectively captured the heavy-tailed, leptokurtic and asymmetric properties of the stocks considered, highlighting its potential to model financial time-series data. In conclusion, our study contributes to the advancement of parameter estimation of MGND models, enhancing the accuracy, reliability, and interpretability of the results. 

\bibliographystyle{apalike}
\bibliography{References.bib}

\paragraph{Disclosure statement} The authors report that there are no competing interests to declare.

\appendix
\section{GND central moments, skewness and kurtosis}
\label{AGND}
The $n$th central moments of the GDN are obtained as follows
\begin{equation}
\begin{split}
E[(X-\mu)^n]&=\frac{\sigma^n\{1+(-1)^n\}\Gamma((n+1)/\nu)}{2\Gamma(1/\nu)}.
\end{split}
\label{eq.GND.centralmoments}
\end{equation}
Mean, variance, skewness and kurtosis obtained from equation \ref{eq.GND.centralmoments} are
\begin{equation}
\begin{split}
&\text{E}(X)=\mu,\\
&\text{VAR}(X)=\frac{\sigma^2\Gamma(3/\nu)}{\Gamma(1/\nu)},\\
&\text{Sk}(X)=0,\\
&\text{Kur}(X)=\frac{\Gamma(1/\nu)\Gamma(5/\nu)}{\Gamma(3/\nu)^2}.
\end{split}
\label{eq.GNDstat}
\end{equation}

\section{MGND central moments, skewness and kurtosis}
\label{AMGND}

The $n$th central moments of the MGND model are obtained as follows
\begin{equation}
\begin{split}
\text{E}\left[X-\text{E}(X)\right]^m=&\sum_{k=1}^K\pi_k\sum_{i=0}^m\binom{m}{i}\Biggr(\mu_k-\sum_{k=1}^K\pi_k\mu_k\Biggr)^{(m-i)}\\
&\times\frac{\sigma_k^i\{1+(-1)^m\}\Gamma((m+1)/\nu_k)}{2\Gamma(1/\nu_k)},
\label{eq.MGND.centralmoments}
\end{split}
\end{equation}
Mean, variance and skewness when $K=2$ are defined as follows
\begin{equation}
\begin{split}
&\text{E}(X)=\pi_1\mu_1+\pi_2\mu_2,\\
&\text{VAR}(X)=\pi_1\Biggr[\mu_1^2+\frac{\sigma_1^2\Gamma{(3/\nu_1)}}{\Gamma{(1/\nu_1)}}\Biggr]+\pi_2\Biggr[\mu_2^2+\frac{\sigma_2^2\Gamma{(3/\nu_2)}}{\Gamma{(1/\nu_2)}}\Biggr]+\pi_1\pi_2(\mu_1-\mu_2)^2,\\
&\text{Sk}(X)=\frac{\pi_1\Bigr[(\mu_1-\pi_1\mu_1)^3+3(\mu_1-\pi_1\mu_1)\frac{\sigma_1^2\Gamma{(3/\nu_1)}}{\Gamma{(1/\nu_1)}}\Bigr]}{\Bigr[(\mu_1-\pi_1\mu_1)^2\frac{\sigma_1^2\Gamma{(3/\nu_1)}}{\Gamma{(1/\nu_1)}}\Bigr]^{\frac{3}{2}}}\\
&\hspace{2cm}+\frac{\pi_k\Bigr[(\mu_2-\pi_2\mu_2)^3+3(\mu_2-\pi_2\mu_2)\frac{\sigma_2^2\Gamma{(3/\nu_2)}}{\Gamma{(1/\nu_2)}}\Bigr]}{\Bigr[(\mu_2-\pi_2\mu_2)^2\frac{\sigma_2^2\Gamma{(3/\nu_2)}}{\Gamma{(1/\nu_2)}}\Bigr]^{\frac{3}{2}}},\\
&\text{Kur}(X)=\frac{\pi_1\Bigr[(\mu_1-\pi_1\mu_1)^3+6(\mu_1-\pi_1\mu_1)^2\frac{\sigma_1^2\Gamma{(3/\nu_1)}}{\Gamma{(1/\nu_1)}}\frac{\sigma_1^4\Gamma{(5/\nu_1)}}{\Gamma{(1/\nu_1)}}\Bigr]}{\Bigr(\Bigr[(\mu_1-\pi_1\mu_1)^2+\frac{\sigma_1^2\Gamma{(3/\nu_1)}}{\Gamma{(1/\nu_1)}}\Bigr]\Bigr)^2}\\
&\hspace{2cm}+\frac{\pi_2\Bigr[(\mu_2-\pi_2\mu_2)^3+6(\mu_2-\pi_2\mu_2)^2\frac{\sigma_2^2\Gamma{(3/\nu_2)}}{\Gamma{(1/\nu_2)}}\frac{\sigma_2^4\Gamma{(5/\nu_2)}}{\Gamma{(1/\nu_2)}}\Bigr]}{\Bigr(\Bigr[(\mu_2-\pi_2\mu_2)^2+\frac{\sigma_2^2\Gamma{(3/\nu_2)}}{\Gamma{(1/\nu_2)}}\Bigr]\Bigr)^2}.
\end{split}
\label{eq.MGNDstatK2}
\end{equation}

\section{Updating equations of the location and shape parameter}\label{up.eqs}
\paragraph{Location parameter}
\begin{equation}
\mu_k^{(m)}=\mu_k^{(m-1)}-\frac{g\bigr(\mu_k^{(m-1)}\bigr)}{g'\bigr(\mu_k^{(m-1)}\bigr)},
\end{equation}
where:
\begin{equation*}\label{eq.muf}
\begin{split}
g\bigr(\mu_k^{(m-1)}\bigr)&=\frac{\nu^{(m-1)}_k}{\bigr(\sigma_k^{(m-1)}\bigr)^{\nu^{(m-1)}_k}}\Biggr(\sum_{x_n\geq \mu_k^{(m-1)}}^Nz_{kn}^{(m-1)}(x_n-\mu_k^{(m-1)})^{\nu^{(m-1)}_k-1}\hspace{5cm}\\
&-\sum_{x_n< \mu_k^{(m-1)}}^Nz_{kn}^{(m-1)}(\mu_k^{(m-1)}-x_n)^{\nu^{(m-1)}_k-1}\Biggr),\\
\end{split}
\end{equation*}

\begin{equation*}\label{eq.mufpartial}
\begin{split}
g'\bigr(\mu_k^{(m-1)}&\bigr)=-\frac{\nu^{(m-1)}_k}{\bigr(\sigma_k^{(m-1)}\bigr)^{\nu^{(m-1)}_k}}\Biggr(\sum_{x_n\geq\mu_k^{(m-1)}}^Nz_{kn}^{(m-1)}(x_n-\mu_k^{(m-1)})^{\nu^{(m-1)}_k-2}\\
&\times(\nu^{(m-1)}_k-1)+\sum_{x_n<\mu_k^{(m-1)}}^Nz_{kn}^{(m-1)}(\mu_k^{(m-1)}-x_n)^{\nu^{(m-1)}_k-2}(\nu^{(m-1)}_k-1)\Biggr).\\
\end{split}
\end{equation*}

After some further calculations, the location parameter of the $k$-\textit{th} component is estimated by the following iteration equation \citep{Bazi2006,Wen2020}
\begin{equation}\label{eq.muk}
\mu_k^{(m)}=\mu_k^{(m-1)}+\frac{A_k}{B_k}
\end{equation}
where
\begin{equation*}
\begin{split}
&A_k=\sum_{x_n\geq \mu_k^{(m-1)}}^Nz_{kn}^{(m-1)}(x_i-\mu^{(m-1)}_k)^{\nu^{(m-1)}_k-1}-\sum_{x_n< \mu_k}^Nz_{kn}^{(m-1)}(\mu^{(m-1)}_k-x_n)^{\nu^{(m-1)}_k-1},\\
&B_k=\sum_{n=1}^Nz_{kn}^{(m-1)}\bigr|x_n-\mu_k^{(m-1)}\bigr|^{\nu_k^{(m-1)}-2}(\nu^{(m-1)}_k-1).
\end{split}
\end{equation*}

\paragraph{Shape parameter}
\begin{equation}
\nu_k^{(m)}=\nu_k^{(m-1)}-\alpha(\nu^{(m-1)})\frac{g\bigr(\nu_k^{(m-1)}\bigr)}{g'\bigr(\nu_k^{(m-1)}\bigr)},
\end{equation}
where:
\begin{equation}\label{eq.nuf}
\begin{split}
g\bigr(\nu_k^{(m-1)}\bigr)=&\sum_{n=1}^Nz_{kn}^{(m-1)}\frac{1}{\nu_k^{(m-1)}}\Biggr(\frac{1}{\nu_k^{(m-1)}}\Psi\Biggr(\frac{1}{\nu_k^{(m-1)}}\Biggr)+1\Biggr)\\
&-\sum_{n=1}^Nz_{kn}^{(m-1)}\Biggr|\frac{x_n-\mu^{(m)}_k}{\sigma^{(m)}_k}\Biggr|^{\nu_k^{(m-1)}} \log \Biggr|\frac{x_n-\mu^{(m)}_k}{\sigma^{(m)}_k}\Biggr|,
\end{split}
\end{equation}
\begin{equation*}\label{eq.nufpartial}
\begin{split}
g'\bigr(\nu_k^{(m-1)}\bigr)=&\sum_{n=1}^Nz_{kn}^{(m-1)}-\frac{1}{\bigr(\nu_k^{(m-1)}\bigr)^{2}}\Biggr(1+\frac{2}{\nu_k^{(m-1)}}\Psi\Biggr(\frac{1}{\nu_k^{(m-1)}}\Biggr)\\
&+\frac{1}{\bigr(\nu_k^{(m-1)}\bigr)^{2}}\Psi'\Biggr(\frac{1}{\nu_k^{(m-1)}}\Biggr)\Biggr)-\sum_{n=1}^Nz_{kn}^{(m-1)}\Biggr|\frac{x_n-\mu^{(m)}_k}{\sigma^{(m)}_k}\Biggr|^{\nu_k^{(m-1)}}\\
&\times\Biggr(\log \Biggr|\frac{x_n-\mu^{(m)}_k}{\sigma^{(m)}_k}\Biggr|\Biggr)^2.
\end{split}
\end{equation*}

The shape parameter of the $k$-\textit{th} component is estimated by the following iteration equation \citep{Bazi2006,Wen2020}
\begin{equation}\label{eq.nuk}
\nu_k^{(m)}=\nu_k^{(m-1)}-\alpha(\nu^{(m-1)})\frac{\sum_{n=1}^Nz_{kn}^{(m-1)}A_k-\sum_{n=1}^Nz_{kn}^{(m-1)}B_k}{\sum_{n=1}^Nz_{kn}^{(m-1)}C_K-\sum_{n=1}^Nz_{kn}^{(m-1)}D_k},
\end{equation}
where
\begin{align*}
&\alpha(\nu^{(m-1)})=e^{-\nu_k^{(m-1)}}\\
&A_k=\frac{1}{\nu_k^{(m-1)}}\Biggr(\frac{1}{\nu_k^{(m-1)}}\Psi\Biggr(\frac{1}{\nu_k^{(m-1)}}\Biggr)+1\Biggr),\\
&B_k=\Biggr|\frac{x_n-\mu^{(m)}_k}{\sigma^{(m)}_k}\Biggr|^{\nu_k^{(m-1)}} \log \Biggr|\frac{x_n-\mu^{(m)}_k}{\sigma^{(m)}_k}\Biggr|,\\
&C_k=-\frac{1}{\bigr(\nu_k^{(m-1)}\bigr)^{2}}\Biggr(1+\frac{2}{\nu_k^{(m-1)}}\Psi\Biggr(\frac{1}{\nu_k^{(m-1)}}\Biggr)+\frac{1}{\bigr(\nu_k^{(m-1)}\bigr)^{2}}\Psi'\Biggr(\frac{1}{\nu_k^{(m-1)}}\Biggr)\Biggr),\\
&D_k=\Biggr|\frac{x_n-\mu^{(m)}_k}{\sigma^{(m)}_k}\Biggr|^{\nu_k^{(m-1)}} \Biggr(\log \Biggr|\frac{x_n-\mu^{(m)}_k}{\sigma^{(m)}_k}\Biggr|\Biggr)^2,
\end{align*}
The digamma $\Psi(1/\nu_k)$ and trigamma $\Psi'(1/\nu)$ functions  are
\begin{equation}
\Psi(1/\nu_k)=\frac{\partial\Gamma(1/\nu_k)}{\partial(1/\nu_k)}\log\Gamma(1/\nu_k),\hspace{0.5cm} \Psi'(1/\nu)=\frac{\partial^2\Gamma(1/\nu_k)}{\partial(1/\nu_k)^2}\log\Gamma(1/\nu_k).
\end{equation}

\section{Algorithms}\label{Algorithms}
\begin{algorithm}[H]
\caption{MGND model estimation via the ECM algorithm.}
\label{algo.MGND_ECM}
\begin{enumerate}
\item \textbf{require}: data $x_1,x_2,...,x_N$
\item \textbf{set the initial estimates}: \textit{k-means} initialization $\text{minimize}\sum_{k=1}^KW(P_k)$, where $P_k$ denotes the set of units belonging to the $k$th cluster and $W(P_k)$ is the within cluster deviance
\item[] \textbf{for} $k=1,\dots,K$ \textbf{do}

$\mu^{(m-1)}_k \leftarrow \text{mean}(P_k)$, $\sigma^{(m-1)}_k \leftarrow \text{std}(P_k)$, $\nu^{(m-1)}_k \leftarrow\text{ randomly generated in }[0.5,3]$, $\pi^{(m-1)}_k \leftarrow\text{ randomly generated in }[0,1]$, $z^{(m-1)}_k \leftarrow$ Eq. \ref{responsibility}

\textbf{end for}

$\log{}L(\theta^{(m-1)}) \leftarrow$ equation \ref{eq.MGNDloglikelihood}, $\epsilon\leftarrow10^{-5}$
\item \textbf{while} $\left[\log{}L(\theta^{(m-1)})-\log{}L(\theta^{(m)})\right] \leq \epsilon$ not convergence \textbf{do} 
\item[] $\mu^{(m)}_k \leftarrow$ Eq. \ref{eq.mu_ls}, $\sigma^{(m)}_k \leftarrow$ Eq. \ref{eq.sigmak}, $\nu^{(m)}_k \leftarrow$ Eq. \ref{eq.inversenu_ls} with $\alpha(\nu^{(m-1)})=1$, $\pi^{(m)}_k \leftarrow$ Eq. \ref{eq.Mweights}, $z^{(m)}_k \leftarrow$ Eq. \ref{responsibility}, $\log{}L(\theta^{(m)}) \leftarrow$ equation \ref{eq.MGNDloglikelihood} 
\item[] \textbf{evaluate} $\left[\log{}L(\theta^{(m-1)})-\log{}L(\theta^{(m)})\right] \leq \epsilon$

\item[] \textbf{for} $k=1,\dots,K$ \textbf{do}

$\pi^{(m-1)}_k \leftarrow \pi^{(m)}_k$, $\mu^{(m-1)}_k \leftarrow \mu^{(m)}_k$, $\sigma^{(m-1)}_k \leftarrow \sigma^{(m)}_k$, $\nu^{(m-1)}_k \leftarrow \nu^{(m)}_k$
\item[] \textbf{end for} 

$\log{}L(\theta^{(m-1)}) \leftarrow \log{}L(\theta^{(m)})$ 
\item[] \textbf{end while} 
\item \textbf{return}: $\theta^{(m)}$, $\log{}L(\theta^{(m)})$
\end{enumerate}
\end{algorithm}

\begin{algorithm}[H]
\caption{MGND model estimation via the ECMs algorithm.}
\label{algo.MGND_GEM}
\begin{enumerate}
\item \textbf{require}: data $x_1,x_2,...,x_N$
\item \textbf{set} $d_k=0$ for $i=1,\dots,K$
\item \textbf{set the initial estimates}: \textit{k-means} initialization $\text{minimize}\sum_{k=1}^KW(P_k)$, where $P_k$ denotes the set of units belonging to the $k$th cluster and $W(P_k)$ is the within cluster deviance
\item[] \textbf{for} $k=1,\dots,K$ \textbf{do} 

$\mu^{(m-1)}_k \leftarrow \text{mean}(P_k)$, $\sigma^{(m-1)}_k \leftarrow \text{std}(P_k)$, $\nu^{(m-1)}_k \leftarrow\text{ randomly generated in }[0.5,3]$, $\pi^{(m-1)}_k \leftarrow\text{ randomly generated in }[0,1]$, $z^{(m-1)}_k \leftarrow$ Eq. \ref{responsibility}

\textbf{end for}

$\log{}L(\theta^{(m-1)}) \leftarrow$ equation \ref{eq.MGNDloglikelihood}, $\epsilon\leftarrow10^{-5}$, $\eta\leftarrow 5^{-3}$

\item \textbf{while} $\left[\log{}L(\theta^{(m-1)})-\log{}L(\theta^{(m)})\right] \leq \epsilon$ not convergence \textbf{do}

\textbf{for} $k=1,\dots,K$ \textbf{do}

$\mu^{(m)}_k \leftarrow$ Eq. \ref{eq.mu_ls}, $\sigma^{(m)}_k \leftarrow$ Eq. \ref{eq.sigmak}, $g(\nu^{(m-1)}_k) \leftarrow Eq. \ref{eq.nuf}$

\textbf{if} $g(\nu^{(m-1)}_k)>\eta$ \textbf{then} $\nu^{(m)}_k \leftarrow$ Eq. \ref{eq.inversenu_ls} with $\alpha(\nu^{(m-1)})=e^{-\nu_k^{(m-1)}}$, $d_k \leftarrow 0$ \textbf{else} $d_k=1$
\textbf{end if} 

$\pi^{(m)}_k \leftarrow$ Eq. \ref{eq.Mweights}, $z^{(m)}_k \leftarrow$ Eq. \ref{responsibility}, 

\textbf{end for}

$\log{}L(\theta^{(m)}) \leftarrow$ equation \ref{eq.MGNDloglikelihood}

\textbf{if} $\sum_{k=1}^K d_k=K$ \textbf{then} 
\textbf{evaluate} $\left[\log{}L(\theta^{(m-1)})-\log{}L(\theta^{(m)})\right] \leq \epsilon$

\textbf{end if}

\item[] \textbf{for} $k=1,\dots,K$ \textbf{do}
\item[] $\pi^{(m-1)}_k \leftarrow \pi^{(m)}_k$, $\mu^{(m-1)}_k \leftarrow \mu^{(m)}_k$, $\sigma^{(m-1)}_k \leftarrow \sigma^{(m)}_k$, $\nu^{(m-1)}_k \leftarrow \nu^{(m)}_k$
\item[] \textbf{end for}

$\log{}L(\theta^{(m-1)}) \leftarrow \log{}L(\theta^{(m)})$ 

\textbf{end while}
\item \textbf{return}: $\theta^{(m)}$, $\log{}L(\theta^{(m)})$
\end{enumerate}
\end{algorithm}

\begin{algorithm}[H]
\caption{Sampling from a MGND with the composition method.}
\label{algo.sim}
\begin{enumerate}
\item \textbf{for} $n \in N$ \textbf{do}
\item[] \textbf{generate} a random $u_n$ from the uniform distribution $U \sim \text{uniform}(0,1)$
\item[] \textbf{if} $\sum_{k=1}^{K-1}\pi_k\leq u_n < \sum_{k=1}^{K} \pi_k$, generate a random $x_n$ from $f_k(x|\mu_k,\sigma_k,\nu_k)$ using the R function \textit{rgnorm} of the package \textit{gnorm}
\item \textbf{return:} simulated data $x_1,x_2,...,x_N$
\end{enumerate}
\end{algorithm}

\section{Tables}\label{Tables}

\begin{table}[H]
\caption{MGND simulations, estimated parameters and RMSE.}
\label{tab.MGNDsim_set1-2}
\centering
\begin{tabular}{lrrrrrrrrr}
\toprule
\multicolumn{9}{c}{Scenario 1}\\
\midrule
&\multicolumn{1}{c}{$\pi_1$}&\multicolumn{1}{c}{$\mu_1$}&
\multicolumn{1}{c}{$\sigma_1$}&\multicolumn{1}{c}{$\nu_1$}&\multicolumn{1}{c}{$\pi_2$}&\multicolumn{1}{c}{$\mu_2$}&\multicolumn{1}{c}{$\sigma_2$}&\multicolumn{1}{c}{$\nu_2$}\\
\multicolumn{1}{c}{$\theta$}&\multicolumn{1}{c}{$0.7$}&\multicolumn{1}{c}{$1$}&
\multicolumn{1}{c}{$3$}&\multicolumn{1}{c}{$5$}&\multicolumn{1}{c}{$0.3$}&\multicolumn{1}{c}{$5$}&\multicolumn{1}{c}{$1$}&\multicolumn{1}{c}{$1.5$}&\multicolumn{1}{c}{$N$}\\
\midrule
\multicolumn{9}{c}{ECM algorithm}\\
$\text{AVG}$ & 0.55 & 0.45 & 2.35 & 5.44 & 0.45 & 4.33 & 1.74 & 2.94 & 250 \\ 
$\text{RMSE}$ & 0.18 & 0.87 & 0.77 & 2.99& 0.18 & 0.97 & 0.93 & 1.96\\
$\text{AVG}$ & 0.58 & 0.49 & 2.48 & 4.51 & 0.42 & 4.49 & 1.66 & 2.47 & 1000 \\ 
$\text{RMSE}$ & 0.15 & 0.59 & 0.59 & 1.02 & 0.15 & 0.60 & 0.80 & 1.25\\  
\multicolumn{9}{c}{ECMs algorithm}\\
$\text{AVG}$ & 0.60 & 0.61 & 2.49 & 4.05 & 0.40 & 4.57 & 1.45 & 2.29 & 250 \\ 
$\text{RMSE}$ & 0.17 & 0.74 & 0.77 & 1.47& 0.17 & 0.84 & 0.82 & 1.28\\  
$\text{AVG}$ & 0.68 & 0.91 & 2.88 & 4.81 & 0.32 & 4.91 & 1.12 & 1.70 & 1000 \\ 
$\text{RMSE}$ & 0.08 & 0.34 & 0.34 & 0.74& 0.08 & 0.33 & 0.41 & 0.61\\ 
\toprule
\multicolumn{9}{c}{Scenario 2}\\
\midrule
&\multicolumn{1}{c}{$\pi_1$}&\multicolumn{1}{c}{$\mu_1$}&
\multicolumn{1}{c}{$\sigma_1$}&\multicolumn{1}{c}{$\nu_1$}&\multicolumn{1}{c}{$\pi_2$}&\multicolumn{1}{c}{$\mu_2$}&\multicolumn{1}{c}{$\sigma_2$}&\multicolumn{1}{c}{$\nu_2$}\\
\multicolumn{1}{c}{$\theta$}&\multicolumn{1}{c}{$0.7$}&\multicolumn{1}{c}{$0$}&
\multicolumn{1}{c}{$1$}&\multicolumn{1}{c}{$5$}&\multicolumn{1}{c}{$0.3$}&\multicolumn{1}{c}{$0$}&\multicolumn{1}{c}{$3$}&\multicolumn{1}{c}{$1.5$}&\multicolumn{1}{c}{$N$}\\
\midrule
\multicolumn{9}{c}{ECM algorithm}\\
$\text{AVG}$ & 0.67 & -0.01 & 1.18 & 13.25 & 0.33 & 0.02 & 3.08 & 1.97 & 250\\ 
$\text{RMSE}$ & 0.14 & 0.16 & 0.89 & 56.65& 0.14 & 0.36 & 1.26 & 1.15\\
$\text{AVG}$ & 0.70 & -0.00 & 1.00 & 5.07 & 0.30 & 0.00 & 3.05 & 1.59 & 1000 \\ 
$\text{RMSE}$ & 0.03 & 0.02 & 0.03 & 0.81 & 0.03 & 0.17 & 0.65 & 0.37\\  
\multicolumn{9}{c}{ECMs algorithm}\\
$\text{AVG}$ & 0.72 & 0.00 & 0.98 & 4.19 & 0.28 & 0.01 & 3.43 & 2.00 & 250 \\ 
$\text{RMSE}$ & 0.06 & 0.06 & 0.07 & 1.35& 0.06 & 0.37 & 1.19 & 0.95\\  
$\text{AVG}$ & 0.71 & -0.00 & 0.99 & 4.65 & 0.29 & 0.00 & 3.22 & 1.67 & 1000 \\ 
$\text{RMSE}$ & 0.03 & 0.02 & 0.03 & 0.71& 0.03 & 0.17 & 0.63 & 0.39\\ 
\bottomrule
\end{tabular}
\end{table}

\begin{table}[H]
\caption{MGND simulations, estimated parameters and RMSE.}
\label{tab.MGNDsim_set3-4}
\centering
\begin{tabular}{lrrrrrrrrr}
\toprule
\multicolumn{9}{c}{Scenario 3}\\
\midrule
&\multicolumn{1}{c}{$\pi_1$}&\multicolumn{1}{c}{$\mu_1$}&
\multicolumn{1}{c}{$\sigma_1$}&\multicolumn{1}{c}{$\nu_1$}&\multicolumn{1}{c}{$\pi_2$}&\multicolumn{1}{c}{$\mu_2$}&\multicolumn{1}{c}{$\sigma_2$}&\multicolumn{1}{c}{$\nu_2$}\\
\multicolumn{1}{c}{$\theta$}&\multicolumn{1}{c}{$0.7$}&\multicolumn{1}{c}{$1$}&
\multicolumn{1}{c}{$1$}&\multicolumn{1}{c}{$2$}&\multicolumn{1}{c}{$0.3$}&\multicolumn{1}{c}{$5$}&\multicolumn{1}{c}{$3$}&\multicolumn{1}{c}{$0.8$}&\multicolumn{1}{c}{$N$}\\
\midrule
\multicolumn{9}{c}{ECM algorithm}\\
$\text{AVG}$ & 0.70 & 1.00 & 1.00 & 2.21& 0.30 & 5.04 & 3.32 & 0.87& 250 \\ 
$\text{RMSE}$& 0.02 & 0.06 & 0.12 & 0.63& 0.02 & 0.53 & 1.43 & 0.24\\
$\text{AVG}$ & 0.70 & 1.00 & 1.00 & 2.06& 0.30 & 5.01 & 3.22 & 0.84& 1000 \\ 
$\text{RMSE}$ & 0.01 & 0.03 & 0.05 & 0.25& 0.01 & 0.25 & 0.75 & 0.11\\  
\multicolumn{9}{c}{ECMs algorithm}\\
$\text{AVG}$ & 0.70 & 1.00 & 0.99 & 2.15& 0.30 & 4.99 & 3.33 & 0.88& 250 \\ 
$\text{RMSE}$  & 0.02 & 0.06 & 0.11 & 0.53  & 0.02 & 0.53 & 1.50 & 0.26\\ 
$\text{AVG}$ & 0.70 & 1.00 & 1.00 & 2.05& 0.30 & 4.99 & 3.15 & 0.83  & 1000 \\ 
$\text{RMSE}$ & 0.01 & 0.03 & 0.05 & 0.25& 0.01 & 0.24 & 0.75 & 0.10\\ 
\toprule
\multicolumn{9}{c}{Scenario 4}\\
\midrule
&\multicolumn{1}{c}{$\pi_1$}&\multicolumn{1}{c}{$\mu_1$}&
\multicolumn{1}{c}{$\sigma_1$}&\multicolumn{1}{c}{$\nu_1$}&\multicolumn{1}{c}{$\pi_2$}&\multicolumn{1}{c}{$\mu_2$}&\multicolumn{1}{c}{$\sigma_2$}&\multicolumn{1}{c}{$\nu_2$}\\
\multicolumn{1}{c}{$\theta$}&\multicolumn{1}{c}{$0.7$}&\multicolumn{1}{c}{$0$}&
\multicolumn{1}{c}{$1$}&\multicolumn{1}{c}{$2$}&\multicolumn{1}{c}{$0.3$}&\multicolumn{1}{c}{$0$}&\multicolumn{1}{c}{$3$}&\multicolumn{1}{c}{$0.8$}&\multicolumn{1}{c}{$N$}\\
\midrule
\multicolumn{9}{c}{ECM algorithm}\\
$\text{AVG}$ & 0.73 & -0.00 & 0.99 & 2.11& 0.27 & 0.04 & 5.62 & 1.28& 250 \\ 
$\text{RMSE}$ & 0.07 & 0.07 & 0.12 & 0.71& 0.07 & 1.00 & 4.18 & 1.07\\
$\text{AVG}$ & 0.72 & -0.00 & 1.00 & 1.99& 0.28 & 0.03 & 4.22 & 0.95 & 1000 \\ 
$\text{RMSE}$ & 0.04 & 0.03 & 0.06 & 0.28& 0.04 & 0.36 & 2.03 & 0.25\\  
\multicolumn{9}{c}{ECMs algorithm}\\
$\text{AVG}$ & 0.73 & -0.00 & 0.99 & 2.02& 0.27 & 0.04 & 5.62 & 1.22& 250 \\ 
$\text{RMSE}$ & 0.07 & 0.07 & 0.12 & 0.52& 0.07 & 0.98 & 4.03 & 0.69\\ 
$\text{AVG}$ & 0.71 & -0.00 & 1.00 & 2.01& 0.29 & 0.02 & 3.93 & 0.92& 1000 \\ 
$\text{RMSE}$ & 0.04 & 0.03 & 0.06 & 0.27& 0.04 & 0.37 & 1.98 & 0.24\\ 
\bottomrule
\end{tabular}
\end{table}

\begin{table}[H]
\caption{Descriptive statistics of daily log-returns on SX5E constituents.}
\label{tab.comp.bs}
\centering
\resizebox{13.5cm}{!}{
\begin{tabular}{lccccccccc}
\toprule
\textbf{Tickers} & \textbf{N} & \textbf{Mean} & \textbf{Median} & \textbf{Std} & \textbf{Skewness} & \textbf{Kurtosis} & \textbf{Min} & \textbf{Max} & \textbf{JB} \\ 
  \midrule
ABI.BE    & 3786 & 0.0126  & 0.0155  & 1.6258 & -0.5261 & 15.2961 & -18.1622 & 13.4532 & 24057  \\
AD.NL     & 3786 & 0.0319  & 0.067   & 1.2615 & -0.3433 & 8.6646  & -10.0198 & 7.7183  & 5145   \\
ADS.DE    & 3749 & 0.0483  & 0       & 1.8658 & 0.2118  & 11.8965 & -16.6886 & 19.3787 & 12409  \\
ADYEN.NL  & 1622 & 0.0684  & 0.1629  & 3.1286 & -1.917  & 50.6746 & -49.4034 & 32.0764 & 155007 \\
AI.FR     & 3786 & 0.0361  & 0.0551  & 1.3019 & -0.1667 & 7.7453  & -11.8337 & 8.1161  & 3576   \\
AIR.FR    & 3786 & 0.0587  & 0.0772  & 2.1305 & -0.3696 & 16.6907 & -25.0623 & 18.6175 & 29692  \\
ALV.DE    & 3749 & 0.0326  & 0.069   & 1.5609 & -0.3385 & 14.1051 & -16.6382 & 14.6728 & 19362  \\
ASML.NL   & 3786 & 0.0902  & 0.1274  & 1.9645 & -0.0342 & 6.535   & -13.1142 & 13.0633 & 1976   \\
BAS.DE    & 3749 & 0.0028  & 0.0496  & 1.6847 & -0.2498 & 6.7933  & -12.5494 & 10.1917 & 2291   \\
BAYN.DE   & 3749 & -0.0189 & 0       & 1.7967 & -0.7681 & 11.7311 & -19.798  & 9.8678  & 12294  \\
BBVA.ES   & 3771 & -0.0071 & 0       & 2.1928 & -0.0765 & 10.1088 & -17.649  & 19.9073 & 7956   \\
BMW.DE    & 3749 & 0.0235  & 0.032   & 1.8212 & -0.3276 & 7.8811  & -13.8933 & 13.5163 & 3795   \\
BN.FR     & 3786 & 0.0109  & 0.0175  & 1.237  & -0.1645 & 7.0022  & -8.8931  & 7.4267  & 2549   \\
BNP.FR    & 3786 & 0.0027  & 0.0407  & 2.2237 & -0.1236 & 11.1418 & -19.1166 & 18.9768 & 10482  \\
CS.FR     & 3786 & 0.0192  & 0.0792  & 1.9485 & -0.165  & 14.6239 & -16.8196 & 19.7782 & 21360  \\
DB1.DE    & 3749 & 0.0359  & 0.0504  & 1.5239 & -0.4082 & 9.8567  & -12.5985 & 12.3142 & 7460   \\
DG.FR     & 3786 & 0.0258  & 0.0498  & 1.6607 & -0.3773 & 15.9178 & -18.7227 & 17.2666 & 26448  \\
DHL.DE    & 3749 & 0.0259  & 0.0701  & 1.5913 & -0.2454 & 7.8797  & -12.8063 & 11.7308 & 3764   \\
DTE.DE    & 3749 & 0.0261  & 0.0105  & 1.3685 & -0.4423 & 9.8393  & -11.2673 & 10.6715 & 7440   \\
EL.FR     & 3786 & 0.0432  & 0.0489  & 1.4374 & 0.0345  & 7.8759  & -10.9702 & 11.1999 & 3758   \\
ENEL.IT   & 3748 & 0.0144  & 0.0323  & 1.6522 & -0.9864 & 14.0645 & -22.1228 & 7.6674  & 19753  \\
ENI.IT    & 3748 & -0.0068 & 0.0602  & 1.6936 & -1.1388 & 21.2244 & -23.3851 & 13.9159 & 52743  \\
IBE.ES    & 3771 & 0.0196  & 0.0444  & 1.4705 & -0.3495 & 11.6154 & -15.1554 & 13.3703 & 11756  \\
IFX.DE    & 3749 & 0.0531  & 0.0625  & 2.2842 & -0.2225 & 6.3665  & -17.0262 & 13.0759 & 1805   \\
INGA.NL   & 3786 & 0.0222  & 0.0373  & 2.3247 & -0.1945 & 12.2608 & -21.5324 & 21.9838 & 13572  \\
ISP.IT    & 3748 & 0.0078  & 0.0565  & 2.405  & -0.6557 & 12.0401 & -26.0594 & 17.962  & 13049  \\
ITX.ES    & 3771 & 0.0468  & 0       & 1.6206 & 0.2489  & 7.5087  & -11.1276 & 13.1323 & 3239   \\
KER.FR    & 3786 & 0.0298  & 0.032   & 1.846  & -0.1225 & 7.4353  & -13.1416 & 10.0708 & 3118   \\
MBG.DE    & 3749 & 0.0169  & 0.0391  & 1.9621 & -0.0801 & 15.3704 & -20.8896 & 24.1193 & 23940  \\
MC.FR     & 3786 & 0.0589  & 0.0719  & 1.7051 & 0.1027  & 6.251   & -9.0777  & 12.0552 & 1677   \\
MUV2.DE   & 3749 & 0.0405  & 0.0852  & 1.4875 & -0.2356 & 21.9209 & -19.5309 & 18.3778 & 56027  \\
NDA.FI.FI & 3695 & 0.0118  & 0.0568  & 1.7814 & -0.4705 & 8.756   & -15      & 12.2009 & 5246   \\
NOKIA.FI  & 3695 & -0.0248 & 0.0433  & 2.4486 & -0.7681 & 20.8655 & -26.5878 & 29.2226 & 49565  \\
OR.FR     & 3786 & 0.0423  & 0.0422  & 1.3715 & 0.1117  & 6.1634  & -7.883   & 8.1003  & 1590   \\
PRX.NL    & 1303 & 0.0126  & -0.0197 & 2.5987 & 0.1068  & 10.5524 & -19.015  & 21.4164 & 3113   \\
RACE.IT   & 2235 & 0.1018  & 0.1098  & 1.7443 & -0.0381 & 8.208   & -10.8247 & 10.4542 & 2534   \\
RI.FR     & 3786 & 0.0194  & 0.035   & 1.3185 & -0.203  & 6.744   & -10.3501 & 7.5608  & 2242   \\
RMS.FR    & 3786 & 0.083   & 0.1071  & 1.5862 & -0.0266 & 8.9923  & -12.5081 & 14.0847 & 5674   \\
SAF.FR    & 3786 & 0.0715  & 0.0492  & 1.997  & -0.5218 & 22.6549 & -25.9726 & 19.0064 & 61187  \\
SAN.ES    & 3771 & -0.0223 & 0.0174  & 2.183  & -0.2228 & 12.5898 & -22.1724 & 20.8774 & 14501  \\
SAN.FR    & 3786 & 0.0159  & 0.0386  & 1.4176 & -1.1407 & 18.5638 & -20.9893 & 6.2345  & 39082  \\
SAP.DE    & 3749 & 0.0476  & 0.0783  & 1.4756 & -1.2883 & 28.0599 & -24.7661 & 11.8219 & 99254  \\
SGO.FR    & 3786 & 0.02    & 0.0168  & 1.9277 & -0.4681 & 10.0025 & -18.7602 & 11.2554 & 7885   \\
SIE.DE    & 3749 & 0.0306  & 0.0484  & 1.6214 & -0.1148 & 7.906   & -13.5774 & 10.9387 & 3774   \\
STLAM.IT  & 3748 & 0.0447  & 0.0826  & 2.5407 & -0.4863 & 7.9363  & -19.6792 & 15.1865 & 3960   \\
SU.FR     & 3786 & 0.0468  & 0.0877  & 1.8194 & -0.1478 & 6.9845  & -15.1032 & 11.3445 & 2523   \\
TTE.FR    & 3786 & 0.0085  & 0.0656  & 1.6322 & -0.5328 & 15.685  & -18.1622 & 14.0407 & 25596  \\
UCG.IT    & 3748 & -0.0163 & 0.0424  & 2.835  & -0.3809 & 9.7444  & -27.1658 & 19.0067 & 7205   \\
VOW3.DE   & 3749 & 0.0142  & 0       & 2.14   & -0.501  & 12.7985 & -22.0877 & 17.434  & 15176  \\
WKL.NL    & 3786 & 0.0606  & 0.0849  & 1.2651 & -0.427  & 7.612   & -10.2898 & 7.6426  & 3476  \\
\bottomrule
\end{tabular}
}
\end{table}

\begin{table}[H]
\caption{AIC of fitted mixture models for daily log-returns on SX5E constituents.}
\label{tab.comp.aic}
\centering
\resizebox{6.5cm}{!}{
\begin{tabular}{lccc}
\toprule
\textbf{Ticker} & \textbf{MGND}     & \textbf{MND}      & \textbf{MSTD}   \\
\midrule
ABI.BE    & \textbf{13447.49} & 13520.69 & 13448.94          \\
AD.NL     & \textbf{11820.63} & 11844.13 & 11823.79          \\
ADS.DE    & \textbf{14640.47} & 14678.05 & 14640.84          \\
ADYEN.NL  & 7711.15           & 8286.08  & \textbf{7708.83}  \\
AI.FR     & 12332.44          & 12355.68 & \textbf{12328.32} \\
AIR.FR    & \textbf{15510.33} & 15619.72 & 15513.31          \\
ALV.DE    & 12937.58          & 13013.83 & \textbf{12934.18} \\
ASML.NL   & \textbf{15466.69} & 15495.17 & 15467.07          \\
BAS.DE    & 14121.55          & 14144.35 & \textbf{14121.49} \\
BAYN.DE   & \textbf{14401.87} & 14491.07 & 14402.2           \\
BBVA.ES   & \textbf{15985.72} & 16044.67 & 15985.89          \\
BMW.DE    & 14613.6           & 14654.23 & \textbf{14611.14} \\
BN.FR     & 11926.2           & 11945.46 & \textbf{11925.55} \\
BNP.FR    & \textbf{15981.38} & 16037.3  & 15984.2           \\
CS.FR     & 14673.27          & 14757.34 & \textbf{14669.39} \\
DB1.DE    & 13232.17          & 13277.43 & \textbf{13229.01} \\
DG.FR     & \textbf{13653.3}  & 13727.49 & 13657.08          \\
DHL.DE    & \textbf{13623.83} & 13661.65 & 13625.99          \\
DTE.DE    & 12273.32          & 12322.28 & \textbf{12269.99} \\
EL.FR     & 12966.39          & 12998.84 & \textbf{12963.41} \\
ENEL.IT   & \textbf{13893.14} & 13980.2  & 13898.66          \\
ENI.IT    & 13825.75          & 13933.24 & \textbf{13817.26} \\
IBE.ES    & 12824.43          & 12875.74 & \textbf{12823.73} \\
IFX.DE    & \textbf{16520.46} & 16554    & 16522.65          \\
INGA.NL   & \textbf{16171.3}  & 16263.78 & 16174.22          \\
ISP.IT    & \textbf{16406.79} & 16488.34 & 16406.79          \\
ITX.ES    & 13948.36          & 13972.2  & \textbf{13947.81} \\
KER.FR    & \textbf{14839.54} & 14863.58 & 14842.25          \\
MBG.DE    & 14961.38          & 15101.6  & \textbf{14959.13} \\
MC.FR     & 14391.26          & 14405.74 & \textbf{14389.41} \\
MUV2.DE   & 12577.93          & 12683.63 & \textbf{12574.16} \\
NDA.FI & \textbf{14121.56} & 14145.87 & 14125.39          \\
NOKIA.FI  & \textbf{15791.62} & 15878.85 & 15795.86          \\
OR.FR     & 12777.22          & 12787.81 & \textbf{12775.61} \\
PRX.NL    & 5979.8            & 6001.5   & \textbf{5978.36}  \\
RACE.IT   & \textbf{8472.49}  & 8486.66  & 8481.12           \\
RI.FR     & 12443.73          & 12456    & \textbf{12442.96} \\
RMS.FR    & 13634.53          & 13672.16 & \textbf{13633.87} \\
SAF.FR    & 14822.74          & 14944.58 & \textbf{14822.04} \\
SAN.ES    & \textbf{15987.48} & 16063.54 & 15989.82          \\
SAN.FR    & 12868.76          & 12918.95 & \textbf{12858.58} \\
SAP.DE    & 12833.25          & 12925.96 & \textbf{12826.25} \\
SGO.FR    & 15146.4           & 15207.78 & \textbf{15144.97} \\
SIE.DE    & \textbf{13793.46} & 13815.34 & 13796.15          \\
STLAM.IT  & \textbf{17117.82} & 17167.35 & 17120.35          \\
SU.FR     & 14854.84          & 14903.31 & \textbf{14852.85} \\
TTE.FR    & \textbf{13711.88} & 13799.63 & 13716.41          \\
UCG.IT    & \textbf{17771.35} & 17819.32 & 17773.8           \\
VOW3.DE   & \textbf{15657.64} & 15776.36 & 15659.83          \\
WKL.NL    & \textbf{12069.45} & 12088.58 & 12070.94          \\ 
\bottomrule
\end{tabular}
}
\end{table}

\begin{table}[H]
\caption{BIC of fitted mixture models for daily log-returns on SX5E constituents.}
\label{tab.comp.bic}
\centering
\resizebox{6.5cm}{!}{
\begin{tabular}{lccc}
\toprule
\textbf{Ticker} & \textbf{MGND}     & \textbf{MND}      & \textbf{MSTD}     \\
\midrule
ABI.BE          & \textbf{13491.16} & 13551.88          & 13492.61          \\
AD.NL           & \textbf{11864.3}  & 11875.33          & 11867.46          \\
ADS.DE          & \textbf{14684.07} & 14709.2           & 14684.45          \\
ADYEN.NL        & 7748.89           & 8313.04           & \textbf{7746.57}  \\
AI.FR           & 12376.11          & 12386.87          & \textbf{12371.99} \\
AIR.FR          & \textbf{15554.01} & 15650.92          & 15556.98          \\
ALV.DE          & 12981.19          & 13044.98          & \textbf{12977.79} \\
ASML.NL         & \textbf{15510.36} & 15526.37          & 15510.74          \\
BAS.DE          & 14165.16          & 14175.49          & \textbf{14165.1}  \\
BAYN.DE         & \textbf{14445.48} & 14522.22          & 14445.8           \\
BBVA.ES         & \textbf{16029.36} & 16075.84          & 16029.54          \\
BMW.DE          & 14657.2           & 14685.38          & \textbf{14654.74} \\
BN.FR           & 11969.88          & 11976.65          & \textbf{11969.22} \\
BNP.FR          & \textbf{16025.05} & 16068.5           & 16027.87          \\
CS.FR           & 14716.94          & 14788.54          & \textbf{14713.07} \\
DB1.DE          & 13275.78          & 13308.58          & \textbf{13272.61} \\
DG.FR           & \textbf{13696.97} & 13758.69          & 13700.75          \\
DHL.DE          & \textbf{13667.43} & 13692.79          & 13669.59          \\
DTE.DE          & 12316.92          & 12353.43          & \textbf{12313.6}  \\
EL.FR           & 13010.06          & 13030.03          & \textbf{13007.08} \\
ENEL.IT         & \textbf{13936.74} & 14011.34          & 13942.26          \\
ENI.IT          & 13869.35          & 13964.38          & \textbf{13860.86} \\
IBE.ES          & 12868.07          & 12906.91          & \textbf{12867.37} \\
IFX.DE          & \textbf{16564.06} & 16585.14          & 16566.25          \\
INGA.NL         & \textbf{16214.97} & 16294.97          & 16217.9           \\
ISP.IT          & \textbf{16450.39} & 16519.48          & 16450.39          \\
ITX.ES          & 13992.01          & 14003.38          & \textbf{13991.45} \\
KER.FR          & \textbf{14883.21} & 14894.78          & 14885.92          \\
MBG.DE          & 15004.98          & 15132.75          & \textbf{15002.74} \\
MC.FR           & 14434.94          & 14436.93          & \textbf{14433.08} \\
MUV2.DE         & 12621.53          & 12714.78          & \textbf{12617.77} \\
NDA.FI       & \textbf{14165.06} & 14176.95          & 14168.89          \\
NOKIA.FI        & \textbf{15835.12} & 15909.93          & 15839.37          \\
OR.FR           & 12820.89          & \textbf{12819.01} & 12819.28          \\
PRX.NL          & 6016              & 6027.36           & \textbf{6014.57}  \\
RACE.IT         & \textbf{8512.47}  & 8515.22           & 8521.1            \\
RI.FR           & 12487.4           & 12487.2           & \textbf{12486.63} \\
RMS.FR          & 13678.21          & 13703.36          & \textbf{13677.54} \\
SAF.FR          & 14866.41          & 14975.77          & \textbf{14865.72} \\
SAN.ES          & \textbf{16031.13} & 16094.71          & 16033.46          \\
SAN.FR          & 12912.43          & 12950.15          & \textbf{12902.25} \\
SAP.DE          & 12876.85          & 12957.1           & \textbf{12869.86} \\
SGO.FR          & 15190.07          & 15238.98          & \textbf{15188.64} \\
SIE.DE          & \textbf{13837.07} & 13846.49          & 13839.76          \\
STLAM.IT        & \textbf{17161.42} & 17198.49          & 17163.95          \\
SU.FR           & 14898.52          & 14934.51          & \textbf{14896.53} \\
TTE.FR          & \textbf{13755.56} & 13830.82          & 13760.08          \\
UCG.IT          & \textbf{17814.96} & 17850.47          & 17817.4           \\
VOW3.DE         & \textbf{15701.25} & 15807.51          & 15703.43          \\
WKL.NL          & \textbf{12113.12} & 12119.77          & 12114.61          \\
\bottomrule
\end{tabular}
}
\end{table}


\end{document}